\documentclass[superscriptaddress,
onecolumn,showpacs,preprintnumbers,amsmath,amssymb]{revtex4}
\usepackage{graphicx}
\usepackage{dcolumn}
\usepackage{bm}
\usepackage{latexsym}

\newcommand{\ct}{\cite}
\newcommand{\bi}{\bibitem}
\newcommand{\be}{\begin{equation}}
\newcommand{\ee}{\end{equation}}
\newcommand{\ba}{\begin{eqnarray}}
\newcommand{\ea}{\end{eqnarray}}

\begin{document}
\title{The effect of long range interactions on the stability of classical
and quantum solids}
\author{Debanjan Chowdhury{\footnote{E-mail: debanjan@iitk.ac.in}}}
\affiliation{Department of Physics, Indian Institute of Technology, Kanpur 208016, India.}
\author{Amit Dutta{\footnote{E-mail: dutta@iitk.ac.in}}}
\affiliation{Department of Physics, Indian Institute of Technology, Kanpur 208016, India.}
\date{\today}%

\begin{abstract}
We generalise the celebrated Peierls' argument to study the stability 
of a long-range interacting classical solid. Long-range interaction
implies that all the atomic oscillators are coupled to each other via
a harmonic potential, though 
the coupling strength decays as a power-law $1/x^{\alpha}$,
where $x$ is the distance between the
oscillators. We show that for the range parameter $\alpha <2$, the long-range
interaction dominates and the one-dimensional system retains a crystalline
order even at a finite temperature whereas for $\alpha \geq2$, the long-range
crystalline order vanishes even at an infinitesimally small temperature.
We also study the effect of quantum fluctuations on the melting behaviour 
of a one-dimensional solid at $T=0$, extending Peierls' arguments to the case of
quantum oscillators.
\end{abstract}
\maketitle
\section{Introduction}

The role of the range of interactions in phase transitions and critical 
phenomena has been investigated for many years. For example, there are 
rigorous theorems which rule out the possibility of long-range ordered 
phases of a system with only short-range interactions \ct{merwag}. It is 
also well known that mean-field theory becomes exact in the limit of infinite 
range interactions \ct{stanley}. Power-law interactions, which fall as 
$1/x^{\alpha}$ with the increasing separation $x$ between the interacting 
elements, can smoothly interpolate between these two extremes 
\ct { anderson70,fisher71}. 
The implications of power-law interactions have been explored for several 
physical systems; these include, for example, classical \ct{kotliar83} 
and quantum spin glasses \ct{dutta02}, ionic systems \ct {pitzer85}, 
complex networks \ct{watts98}, etc. Very recently, non-equilibrium phase 
transitions with long-range interactions have also received attention
\ct{hinrichsen07}. 

Power-law interactions can lead to non-trivial and counterintuitive 
results. For example, for the $d$-dimensional Ising model with 
ferromagnetic power-law interactions, the critical exponents depend
on the range parameter $\alpha$  and the conventional short-range 
critical exponents are recovered for relatively large $\alpha$
(typically for $\alpha \to 2$) \ct{fisher71, jenssen97}. 
The ferromagnetic to paramagnetic transition in a $d$-dimensional 
Ising model with inverse-square interaction $1/x^2$ is a special 
type of Kosterlitz-Thouless transition \ct{kosterlitz} driven by 
topological defects \ct{kosterlitz76}.

For many systems, heuristic reasoning, which are essentially extensions 
of Peierls' original arguments \ct{peierls}, have been remarkably 
successful in predicting the (im-)possibility of long-range order. 
Some of these results were later established using rigorous analytical 
and numerical methods. Examples include predicitions of the 
lower-critical dimension of the Ising model and the impossibility
of the existence of a two-dimensional crystaline order at a finite 
temperature. 

In this work, employing Peierls'-type arguments, we examine the stability 
of long-range crystalline order in a d-dimensional system with 
power-law interactions. In other words, we consider a crystal at a
finite temperature where the atomic oscillators interact via a harmonic 
potential with a strength that decays algebraically, i.e., as 
$1/x^{\alpha}$, with $x$ being the seperation between the oscillators.
It is to be noted that, the power-law form of interaction does not 
have an inherent length scale associated with it. This is in sharp 
contrast to the case where the strength of the interaction between 
two oscillators separated by a distance $x$ decays exponentially as
 $e^{-x/L_0}$.  Clearly, in the latter case, $L_0$ sets in a length 
scale over which the interactions are effective and therefore the 
melting behaviour turns out to  be identical to oscillators with 
only nearest neighbour interactions.

The Peierls' argument generalised to the power-law interactions 
suggests that the crystalline order persists for a one-dimensional 
solid even at a finite temperature, provided the range parameter 
$\alpha <2$ whereas for $\alpha\geq 2$ the crystalline order is 
destroyed by infinitesimal thermal fluctuations. 
The case $\alpha =2$ corresponds to a marginal situation. We also comment
on the stability of the crystalline structure at zero-temperature when
the oscillators are quantum mechanical so that quantum fluctuations 
tend to destroy the crystalline order.

\section{Long Range Interactions}
\begin{figure} 
\begin{center}
\includegraphics[width=0.75\columnwidth]{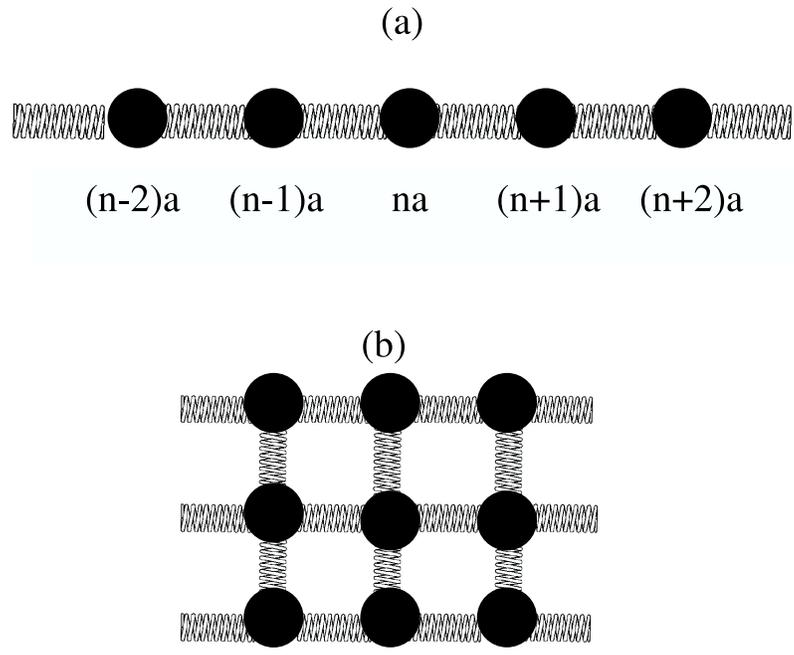}
\end{center}
\caption{(a) A One dimensional crystal and a (b) Two dimensional crystal.} 
\label{crystals}
\end{figure}
Let us consider a one dimensional crystalline arrangement of harmonic 
oscillators 
with long-range interactions in the absence of any thermal or quantum 
fluctuations. The potential energy for this system(Fig.{\ref{crystals}}) 
in the harmonic approximation  
can be expressed as {\cite{mermin}},
\begin{equation}
U={\sum_{n}}{\sum_{m}}{\dfrac{K_{m}}{2}}[q(na)-q([n+m]a)]^{2}
\label{lrpot}
\end{equation}
where $K_m$ is the effective spring constant, 
$a$ is the interatomic seperation 
and $n$ is the index representing the discrete
sites on the crystal. The equation of motion for the atom at the 
position $`na'$ is given by
\begin{equation}
M{\ddot{q}}(na)={\sum_{m>0}}K_{m}[q([n+m]a)+q([n-m]a)-2q(na)],
\label{eqnlr}
\end{equation}
where $M$ is the mass of an oscillator. 
We assume the usual plane wave solutions of the type $q(na)\propto e^{i(kna-wt)}$, 
to obtain the  dispersion relation of the many-oscillator system
given as
\begin{equation}
{\omega}=2{\sqrt{\dfrac{{\sum_{m>0}}K_{m}sin^{2}(mka/2)}{M}}}
\label{displr}
\end{equation}
In the long wavelength limit, i.e. $k$$\rightarrow$0, which
is relevant in determining the phase transition behaviour,
 Eqn.(\ref{displr}) reduces to,
\begin{equation}
{\omega}=a {\sqrt{\dfrac{{{\sum}_{m>0}}K_{m}m^{2}}{M}}} |k|,
\label{displrn}
\end{equation}
provided $\sum_{m>0}m^2K_m$ converges. Now in the  special case of
 nearest neighbour short range interactions,  
we recover the standard phononic \ct{kittel} dispersion relation, 
${\omega}=ck$, where $c$ is the speed of sound. 
In the case of long range interactions, on the other hand,  the interaction 
strength decays  with distance 
as $K_m=\dfrac{1}{m^{\alpha }}$ ($\alpha >$0), then ${\omega}{\sim}|k|$ holds for $\alpha\ge4$ since the sum in Eqn.({\ref{displrn}}) 
converges. For $1<\alpha<3$, we get (see Appendix I) 
\begin{equation}
 {\omega}{\sim}k^{(\alpha-1)/2}.
\label{lrdispsqrt}
\end{equation}
It can be shown that the above dispersion relation holds for higher 
dimensional cases.  
We shall restrict our discussion to the range
$1<\alpha\leq 3$ to
estimate the marginal dimension or the range of the stability of a crystalline
structure.

We shall now examine the effect of thermal fluctuations on 
the stability of the above solid using the formalism
developed by Peierls' {\cite{peierls}}. In the real space, 
the displacement of the atom 
at the position $`na'$ can be written  as a superposition of 
the different Fourier modes as,
\begin{equation}
q(na)={\sum}_{k} q_{k} e^{ikna},
\label{founa}
\end{equation}
$0<k<2\pi/a$, where the time dependence 
has been ignored. 
Making use of the classical 
equipartition theorem, we get
\begin{equation}
{\langle}|q_{k}|^{2}{\rangle}={\langle}q_{k}~q_{-k}{\rangle}=
{\dfrac{k_{B}T}{Mw_{k}^{2}}},
\label{equic}
\end{equation}
The real space 
displacement, $q(na)$ is immediately found to be,
\begin{equation}
{\langle}q^{2}(na){\rangle}={\int} {\langle}q_{k}~q_{-k}{\rangle} e^{ikna} d^{d}k={\dfrac{k_{B}T}{M}}{\int}{\dfrac{e^{ikna}}{w_{k}^{2}}} d^{d}k,
\label{realamp1}
\end{equation}
where $d$ denotes the dimensionality of space.
Then $\langle$$\delta$$^2$$\rangle$=$\langle$q$^2$$(na)$$\rangle$
gives a measure of the fluctuation in the position of the atom at $`na'$. 
To probe the marginality in the melting behaviour, we shall focus on
 the fluctuations in the displacement of  the atom at the origin  ($na$=0). 

Let us first enlist the form of $\langle$$\delta$$^2$$\rangle$ for 
solids with short range interactions in different spatial dimensions
\begin{eqnarray}
{\langle}{\delta}^{2}{\rangle}~
\begin{cases} {\sim}{{\int}_{0}^{2{\pi}/a}}{\dfrac{dk}{k^{2}}} ~~{\rm for~ d=1}, \\  \\
              {\sim}{{\int}_{0}^{2{\pi}/a}}{\dfrac{k dk}{k^{2}}} ~~{\rm for~ d=2}, \\  \\
              {\sim}{{\int}_{0}^{2{\pi}/a}}{\dfrac{k^{2} dk}{k^{2}}}~~{\rm for~ d=3},
\end{cases}
\label{del2sr}
\end{eqnarray} 
From the above Eqn.({\ref{del2sr}}), 
we see that  $\langle$$\delta$$^2$$\rangle$ diverges 
for a one and two dimensional solid in the limit $k\to 0$.
 Moreover, the 2-d solid shows a $\log$ 
divergence, which means that the divergence is  extremely slow. This 
log-divergence immediately establishes the spatial dimension $d=2$ as the 
marginal dimension. For $d=3$, on the other hand, $\langle$$\delta$$^2$$\rangle$ is finite in the limit $k\to 0$, pointing to the existence of 
the crystalline order even at a finite 
temperature in a three-dimensional solid. 

Let us  now extend  the above  analysis to solids with long range 
interactions. The form of the fluctation $\langle$$\delta$$^2$$\rangle$ is
enlisted below:
\begin{eqnarray}
{\langle}{\delta}^{2}{\rangle}~
\begin{cases} {\sim}{{\int}_{0}^{2{\pi}/a}}{\dfrac{dk}{k^{\alpha-1}}}{\sim}{\dfrac{1}{k^{\alpha-2}}} ~~{\rm for~ d=1}, \\ \\
              {\sim}{{\int}_{0}^{2{\pi}/a}}{\dfrac{dk}{k^{\alpha-2}}}{\sim}{\dfrac{1}{k^{\alpha-3}}} ~~{\rm for~ d=2}, \\  \\
              {\sim}{{\int}_{0}^{2{\pi}/a}}{\dfrac{dk}{k^{\alpha-3}}}{\sim}{\dfrac{1}{k^{\alpha-4}}}~~{\rm for~ d=3},\\ \\
               1<\alpha<3 
\end{cases}
\label{del2lr}
\end{eqnarray}
The case with $d=1$ turns out to be the most interesting as
 we observe that for all values  $ \alpha < 2$, the crystal remains stable even at a finite temperature. At $\alpha=2$, we observe the first signature of a 
divergence. Therefore, we conclude that $\alpha =2$ happens to be the marginal 
case, with 
fluctuations diverging logarithmically with the system size. This settles 
$\alpha=2$ as the marginal case for a one-dimensional solid. If the long range
interaction decays faster than $1/x^{2}$, the interaction is feeble and
the solid fails to retain the crystal structure even at an infinitesimally 
small temperature.  

\section{T=0 : Effects of Quantum Fluctuations}
In Eqn.({\ref{realamp1}}), if we put temperature  $T=0$, the average 
fluctuation $\langle$$\delta$$^2$$\rangle$ vanishes.  
We now have to take care of the quantum fluctuations if the harmonic oscillators
are quantum mechanical.
In that case,  we should now set the average energy of a mode 
to be the zero-point energy $\hbar \omega_k$ of the corresponding
oscillator. We obtain
\begin{eqnarray}
{\langle}q_{k}~q_{-k}{\rangle}&=&{\dfrac{{\hbar}w_{k}}{Mw_{k}^{2}}}={\dfrac{{\hbar}}{Mw_{k}}},\\
{\langle}q^{2}(na){\rangle}&=&{\int} {\langle}q_{k}~q_{-k}{\rangle} e^{ikna} d^{d}k={\dfrac{\hbar}{M}}{\int}{\dfrac{e^{ikna}}{w_{k}}} d^{d}k \nonumber. \\
\label{equiq}
\end{eqnarray}
The above relations lead to a non-trivial modification in the marginal 
dimension of a solid with quantum mechanical oscillators, as illustrated below.
In a spirit similar to that of the previous section, let us now examine
 the $\langle$$\delta$$^2$$\rangle$ for a quantum solid
 with short range interactions,
\begin{eqnarray}
{\langle}{\delta}^{2}{\rangle}~
\begin{cases} {\sim}{{\int}_{0}^{2{\pi}/a}}{\dfrac{dk}{k}} ~~{\rm for~ d=1}, \\
               \\  \\
              {\sim}{{\int}_{0}^{2{\pi}/a}}{\dfrac{k dk}{k}} ~~{\rm for~ d=2}, \\  \\
              {\sim}{{\int}_{0}^{2{\pi}/a}}{\dfrac{k^{2} dk}{k}}~~{\rm for~ d=3},
\end{cases}
\label{del2srq}
\end{eqnarray}
Interestingly, under the effect of quantum fluctuations, 
both two and three dimensional crystals sustain the long range crystalline order, whereas
the one dimensional solid shows a logarithmic divergence in the limit $k\to 0$. Hence in the case of quantum oscillators at $T=0$, 
$d=1$ is the marginal dimension whereas in higher dimensions
a perfect crystalline order exists at absolute zero. The above finding when
contrasted with the melting behaviour of a classical solid with short-range
interacting oscillators, an interesting $d_{quantum}\to (d+1)_{classical}$ 
correspondence, with respect to the marginal dimensionality arises.

Let us now assume that the quantum oscillators are also long-range interacting 
as
the classical oscillators of the previous section. 
A similar line of arguments leads to 
\begin{eqnarray}
{\langle}{\delta}^{2}{\rangle}~
\begin{cases} {\sim}{{\int}_{0}^{2{\pi}/a}}{\dfrac{dk}{k^{(\alpha-1)/2}}}{\sim}{\dfrac{1}{k^{(\alpha-3)/2}}} ~~{\rm for~ d=1}, \\ \\
              {\sim}{{\int}_{0}^{2{\pi}/a}}{\dfrac{dk}{k^{(\alpha-3)/2}}}{\sim}{\dfrac{1}{k^{(\alpha-5)/2}}} ~~{\rm for~ d=2}, \\  \\
              {\sim}{{\int}_{0}^{2{\pi}/a}}{\dfrac{dk}{k^{(\alpha-5)/2}}}{\sim}{\dfrac{1}{k^{(\alpha-7)/2}}}~~{\rm for~ d=3},\\ \\
               1<\alpha<3
\end{cases}
\label{del2lrq}
\end{eqnarray}
It is obvious  from the above expressions that the question of any instability
does not arise even in $d=1$ if $\alpha < 3$. Therefore a one dimensional
long-range quantum solid retains the
crystalline order for a finite strength of quantum fluctuations, if  
the decay of the long-range interactions with the distance between the oscillators
is sufficently slow. 
When $\alpha=3$,
\begin{eqnarray}
{\langle}{\delta}^{2}{\rangle}{\sim}{{\int}_{0}^{2{\pi}/a}}{\dfrac{dk}{k {\sqrt{\ln k}}}},\\
{\langle}{\delta}^{2}{\rangle}{\sim}{\int}{\dfrac{dt}{\sqrt{t}}}{\sim}{\sqrt{t}}{\sim}{\sqrt{\ln k}},
\label{alpha3quant1d}
\end{eqnarray}
which corresponds to the first sign of a divergence as $\alpha\to3$, and hence 
the situation is inferred to be marginal. The strength of the divergence keeps 
increasing till $\alpha=4$, when the $\langle$$\delta$$^{2}$$\rangle$$\sim$$\ln k$.

We have summarised the different stability criteria obtained till now in
Table I.   
\begin{table}
\caption{Stability of Solids}
\begin{tabular}{|l||l|l||l|l|}
\hline
 &\multicolumn{2}{l|}{Short range interactions}&\multicolumn{2}{l|}{Long Range Interactions}\\
\cline{2-5}
Spatial&Classical&Quantum&Classical&Quantum\\
dimension &(T${\ne}$0)&(T=0)&(T${\ne}$0)&(T=0)\\
\hline\hline
d=1&Unstable&Unstable&Stable(1$<$$\alpha$$<$2)&Stable(1$<$$\alpha$$<$3)\\
& & ({$\ln$}) &Unstable($\alpha${$\geq$}2)&Unstable($\ln$)\\
 &&&&for ($\alpha$$\ge$3)\\
d=2&Unstable &Stable&Stable(1$<$$\alpha$$<$3)&Stable\\
& (${\ln}$)&&Unstable($\ln$) &\\
&&&for $\alpha$$\ge$3&\\
d=3&Stable&stable&Stable&Stable\\
\\
\hline
\end{tabular}
\label{table2}
\end{table}
\section{ Conclusion}
In this work, we have analyzed the stability of classical and
quantum  solids using Peierls' argument, with a special emphasis on the
one dimensional situation. We just note in the passing that mercury chain salts, in which mercury is intercalated into linear
chains in $AsF_6$ to give a compound $Hg_{3-\delta}AsF_6$ ($\delta\sim$0.18 at 300K), is a system that closely resembles
a one dimensional solid {\cite{chaikin}}.
                                                                                
In the presence of short range interactions, we never expect the 1-d 
solid to exist at any finite temperature
(Eqn.{\ref{del2sr}}). An infinitesimal thermal disturbance is sufficient to destroy
 long range order.
The existence of long range interactions among the atomic 
oscillators, where the interaction strength falls off with
distance as $1/x^{\alpha}$ with $\alpha <2$, has a non-trivial effect on the
stability of the one-dimensional crystal. Using a variant of Peierls' argument,
we have shown that even a one-dimensional  solid can sustain the 
crystalline order at a finite
temperature  if the range-parameter $\alpha$ is sufficiently small ($\alpha <2$).   For $\alpha =2$,  $\langle$$\delta$$^2$$\rangle$ shows a logarithmic 
divergence  and hence this
is the marginal case of stability. This is similar to what is observed 
 also in the case of a ferromagnetic power-law interacting Ising model {\cite{kosterlitz76}}.That 
the inverse-square interaction is the marginal dimension even for melting 
of a long-range solid is an interesting observation. It should be mentioned that
the situation $\alpha=2$ turns out to be a marginal situation in various other
long range interacting systems, for example in fracture models with variable 
range interactions \ct{herrman} or in networks with long range links 
\ct{moukarzel}.

We have also explored the marginal dimension in the case where the oscillators 
are quantum mechanical. Using a generalised  version of  Peierls' argument
 extended to
the quantum case, which to our knowledge is new,
 we observe an interesting $d_{quantum} \to (d+1)_{classical}$ 
correspondence (Table {\ref{table2}}). 
A two-dimensional short-range classical solid is marginally stable against 
thermal fluctuations
whereas a one-dimensional short range quantum solid happens to be marginally stable against
quantum fluctuations at $T=0$. 
This reminds us of the well known correspondence between the critical behaviour
of a $d$-dimensional quantum Ising model and $(d+1)$-dimensional classical Ising
model \ct{sachdev}. However, at an infinitesimally small temperature a one-dimensional 
quantum solid can not retain a long-range order, which means that the 
quantum effects are irrelevant at a finite temperature as the zero-point 
energy is vanishingly small compared to the thermal energy $k_B T$. A unique
correspondence emerges out of our study. We have shown that for a long-range
one-dimensional classical solid, $\alpha=2$ is the marginal range of interaction. In the corresponding quantum case, $\alpha =3$ turns out to be the marginal 
case. 
There seems to exist a quantum-classical correspondence in the marginal range ofinteraction, 
just as in the case of dimension.

One may also wonder about the existence of a tunable parameter in a 
quantum solid which may be varied to change the strength of the quantum fluctuations. 
 Eqn.({\ref{equic}}), suggests that the mass of the atomic oscillator is
an appropriate candidate for the above as the more massive the
atom, the less are the fluctuations or uncertainty in its position.

We should conclude with the comment that in deriving the marginal dimensions
or ranges in above mentioned solids, we have  used the Peierls' argument and
its generalisation to the quantum case. Although, Peierls argument is 
fairly successful in predicting the marginal dimension or range in several
other situations, it fails to throw any light on the transition temperature
(for dimensions above the marginal dimension), nature of transitions or
role of defects in the transition. However as we have shown, it predicts
interesting results even in the case of melting of classical
and quantum solids.

\section{Acknowledgements}
We thank J.K. Bhattacharjee for fruitful discussions. AD acknowledges 
useful discussions and collaborations with S. M. Bhattacharjee
and B. K. Chakrabarti in related fields. This work is partially 
supported by a KVPY fellowship to DC.  
 
\section{Appendix {\rm I}}
When the sum in Eqn.({\ref{displrn}}) is divergent, we recast Eqn.({\ref{displr}}) in the form of an integral as,
\begin{equation}
{\omega}=2{\sqrt{\dfrac{{\int_{0}^{\infty}}K_{m}sin^{2}({\beta}m) dm}{M}}},
\label{int1}
\end{equation}
where $\beta$=$ka/2$. Now the integral is evaluated as follows($1<\alpha<3$),
\begin{equation}
{\int_{0}^{\infty}}{\dfrac{sin^{2}({\beta}m) dm}{m^{\alpha}}}={\beta}^{\alpha-1}{\int_{0}^{\infty}}{\dfrac{sin^{2}(y) dy}{y^{\alpha}}}
\label{int2}
\end{equation}
Since the second integral is just a constant, we obtain Eqn.({\ref{lrdispsqrt}}).
For the special case $\alpha$=3, the integrand in Eqn.({\ref{int2}}) diverges 
at $m=0$. In this case, the dipersion relation turns out to be \ct{mermin}, 
$\omega$=$k{\sqrt{|\ln~k|}}$. 

\end{document}